\documentclass[review]{elsarticle}

\usepackage{lineno,hyperref}
\modulolinenumbers[5]

\journal{Journal of \LaTeX\ Templates}

\begin{document}
\begin{frontmatter}

\title{Multiparticle collision dynamics in porous media}


\author{Maciej Matyka}
\address{Faculty of Physics and Astronomy, University of Wroc{\l}aw, pl. M. Borna 9, 50-204 Wroc{\l}aw, Poland}
\ead{maciej.matyka@uwr.edu.pl}


\begin{abstract}
Multiparticle collision dynamics (MPCD) is a relatively new algorithm of fluid flow simulations that has been applied mostly to flows around simple objects. One might ask how it behaves in more complex flows. Therefore, we extend MPCD to simulate transport in porous media.
For this, a particle-level drag force is introduced into the original algorithm. The force hinders the flow, which results in global resistance to flow and decrease of permeability. The extended algorithm is validated in the flow through a porous channel and compared with an analytical solution. Some basic properties of the solver are investigated.
\end{abstract}


\end{frontmatter}

\linenumbers

\section{Introduction}

Porous media flows are ubiquitous in nature. An example is a flow through porous sediments that occur at the ocean bottom, where sandy sediment is exposed to bottom current flows \cite{Huettel96,Guo02}. There, the transport of the water in porous media is important for many natural processes, including CO$_2$ storage and denitrification \cite{Gao12}. In-situ experiments in such systems require the usage of complex experimental techniques \cite{Lin15}, thus, numerical simulations are often the best option. 

Numerically, porous media are modelled either at the microscopic or macroscopic level. In microscale the porous matrix exists explicitly as a system of interconnected void areas (pores) between solids. In this description the system size is limited mostly by the computer memory required for the numerical mesh.
In the macroscale, the Navier-Stokes equation may be extended by a resistant term \cite{Guo02} to simulate effect of porous media.

In general, fluids are often modelled by integration of the Navier-Stokes equations discretized using numerical grids. There are many numerical methods of this type and they require a tedious task of numerical mesh generation that is limited by both the acceptable time and available computational resources, especially in complex geometries.
Another option is to describe the system at the mesoscopic level with the Lattice Boltzmann Method velocity distribution function defined on a regular grid. One may also use particles that are driven by the flow. An example is the multiparticle collision dynamics (MPCD). This method, originally used for systems that include thermal fluctuations, may be used for standard hydrodynamics as well \cite{Angelis12}. Recently, MPCD was used in hydrodynamical applications, see e.g. vortex shedding in the flow past cylinder \cite{Malevanets99}, flow over fish-like objects \cite{Reid09}, slip and no-slip steady flow past cylinders \cite{Bedkihal2013}. The advantage of MPCD over standard solvers is a relative simplicity of its implementation and no need to use a numerical mesh.

The aim of this work is to utilize advantages of MPCD and extend the original algorithm so as it can be used to simulate flows in macroscopic porous media (without explicit geometry of pores). We show that after extension of the standard method with velocity rescaling procedure we are able to obtain correct  numerical profiles in a porous channel. The numerical results are convergent with decreasing time step in the model. We show that Darcy's Law holds in the model and test the basic numerical properties of the model.

\section{The model}

In MPCD the fluid is represented by point particles. The simulations consists of two steps: streaming and collision. 
No explicit particle-particle interaction potential exists, which distinguishes the model from other particle-based methods such like smoothed particle hydrodynamics and molecular dynamics.

In the streaming step, the $i$-th particle moves at a constant velocity:
\begin{equation}
\vec{r}_i(t+h)=\vec{r}_i(t)+h\vec{v}_i(t),
\end{equation}
where $\vec{r}_i(t)$ is its position at time $t$, $\vec{v}_i$ is its velocity assumed to be constant at the time interval $h$.

\begin{figure}[!ht]
\centering
\includegraphics[width=0.5\columnwidth]{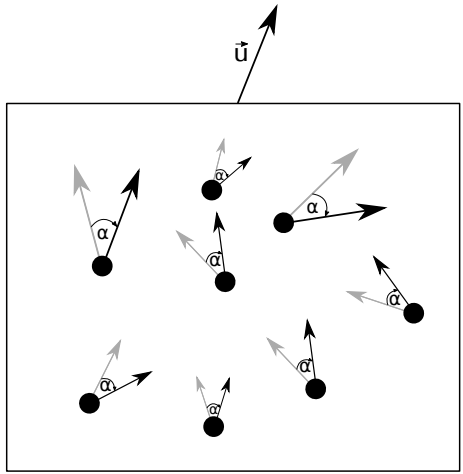}
\caption{Scheme of MPC collision step. Rotation of velocities of particles grouped in a single cell in the moving frame of reference. The velocity of the frame $\vec{u}$ is an average taken over all particles in the cell. Vectors before rotation are drawn with light gray. All particles are rotated by the same, random angle $\alpha$.\label{fig:rotate}}
\end{figure}

To perform the collision step particles are grouped into a regular grid. Grid resolution is adjusted, so that on average there are $70$ particles in a single cell (particle positions are initialized as random). The velocity of the $i$-th particle is rotated in the frame of reference moving with the average velocity $\vec{u}$ computed as the average over all its neigbours in the cell (see Fig.~\ref{fig:rotate}):
\begin{equation}\label{eq:rotation}
\vec{v}_i(t+h)=\vec{u}+s\mathbf{R}(\alpha)(\vec{v}_i-\vec{u}),
\end{equation}
where $\mathbf{R}(\alpha)$ is the rotation matrix by the angle $\alpha$ and $s$ is the scaling variable representing a thermostat (which we calculate individually for each cell). For MPCD particles of equal unit mass we use the following expression:
\begin{equation} \label{eq:thermostat}
s=\sqrt{\frac{k_{B}Td(n-1)}{\sum_{i=1}^n
|\vec{v}_i-\vec{u}|^2}},
\end{equation}
where $k_B$ is the Boltzmann constant, $T$ is the temperature, $d$ is the number of dimensions, $n$ is the number of particles in a single cell, and the sum goes over all particles in the cell \cite{Bolintineanu12,Wysocki10}. 

To ensure the Galilean invariance of the model, the cell borders are shifted randomly after each computational step \cite{Wysocki10}. For this, at each step, we draw two random numbers $r_x\in(-r/2,r/2)$ and $r_y\in(-r/2,r/2)$ and use the shifted coordinates of cells in the grouping procedure. Here $r$ denotes the width and height of a cell in the grid; thus they are shifted by at most half of their linear size in each direction.

\subsection{Porous media}

To simulate effect of porous media on the flow, we use a similar approach to that reported in \cite{Dardis98}. There, velocity of the fluid was locally damped. The difference is, however, that in contrast to \cite{Dardis98}, where the velocity distribution function is used, our model works on particles and their velocities.

To model porous media we assume that the velocity of the fluid in the pore-space is low and the Reynolds number$Re<1$. In such scenario the following drag force acts on each particle:
\begin{equation}\label{eq:drag}
\vec{f}_i = -C\vec{v}_i,
\end{equation}
where $C$ is the drag coefficient - some function of porosity, geometry and other parameters of the porous matrix. The discretized equation for the momentum change reads:
\begin{equation}\label{eq:scaling}
\vec{v}_i = \vec{v}_i(1 - hC)=\lambda \vec{v}_i,
\end{equation}
where $h$ is the discretization time step. Thus, to include drag, we have to rescale the velocity by a single factor $\lambda=1-hC$. The time step must be small enough to ensure that $0<\lambda<1$, i.e.
$0<h<1/C$.
The procedure is applied to all particles that flow in the porous domain at each simulation step. In general, the drag coefficient $C$ may be space and/or time dependent ($C=C(\vec{r},t)$), which leads to many applications, including coupled fluid-porous and porous-porous systems with a variable porosity (not considered in the paper).

\section{Results and discussion}

The velocity scaling described with Eq.~(\ref{eq:scaling}) modifies the permeability of the modelled medium. To investigate properties of the permeability, I simulated the fluid flow in a straight, 2D rectangular channel filled with a homogeneous porous medium. Such a system may be solved analytically with the damped Stokes model \cite{Dardis98,Balasubramanian87}:
\begin{equation}
\nu\frac{d^2u}{dx^2}-\alpha u=\frac{1}{\varrho}\frac{dp}{dy}.
\end{equation}
The exact solution reads \cite{Dardis98}:
\begin{equation}
\label{eq:solution}
u(x)=-\frac{1}{\varrho\alpha}\frac{dp}{dy}\left[1-\frac{\cosh\left[r(x-W/2)\right]}{\cosh(rW/2)}
\right],
\end{equation}
where $x$ is the horizontal position in the channel, $\alpha$ is a parameter related to the damping force, $r$ is a free parameter, $\frac{dp}{dy}$ is the pressure gradient along the vertical direction, and $W$ is the channel width.

We started with a flow through a fully porous channel. We used the no-slip condition at the top and bottom walls. The vertical boundary was periodic. The gravity was applied to simulate the flow in the vertical direction.
We used a regular grid of $64\times 64$ cells in the collision step. On average, $70$ particles are initially placed in each cell at random positions. In the collision step $\alpha=\pm 160^o$ was used. In the thermostat, Eq.~(\ref{eq:thermostat}), we kept $k_BT=0.01$ \cite{Lamura01}.
To verify our solutions, we used the least square fitting of Eq.~(\ref{eq:solution}) to simulation data.

Our first test was to check the minimum time necessary for the solution to converge to the stationary state. To this end we measured the total kinetic energy in our simulation. Because the flow was driven by gravity, we expected it to converge to the stationary state. The resulting plots revealed that at least $10000$ time steps are required for $h=0.001$ to achieve the steady solutions (data not shown).
Thus, in all our simulations the total number of $10 000$ time steps was used in each case. Moreover, we averaged the macroscopic profiles of the velocity 
starting from step $5000$ with the $n=200$ interval.

\begin{figure}[!ht]
\centering
\includegraphics[width=0.7\columnwidth]{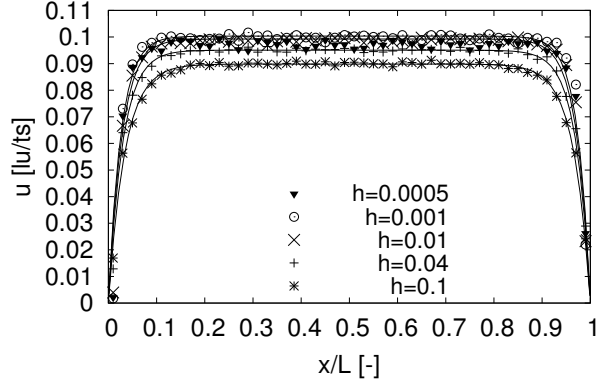}
\caption{Convergence test of the velocity profiles in a fully porous channel for time steps $h=0.1, 0.05, 0.01, 0.001$ and $0.0005$. Data points represent our MPCD simulation results, solid lines are the numerical fits of Eq.~\ref{eq:solution}.\label{ploth}}
\end{figure}

In the next numerical test we verified convergence of the numerical solutions with a decreasing time step $h$. We simulated the flow in the channel at a fixed drag coefficient $C=1$. Thus, the scaling factor $\lambda(h)$ varied with $h$ according to Eq.~(\ref{eq:scaling}). The resulting velocity profiles at varying $h$ are plotted in Fig.~\ref{ploth}. From these data the minimum useful time step was estimated as $h=0.01$ and we used this value in the remaining part of the study.

\begin{figure}[!ht]
\centering
\includegraphics[width=0.7\columnwidth]{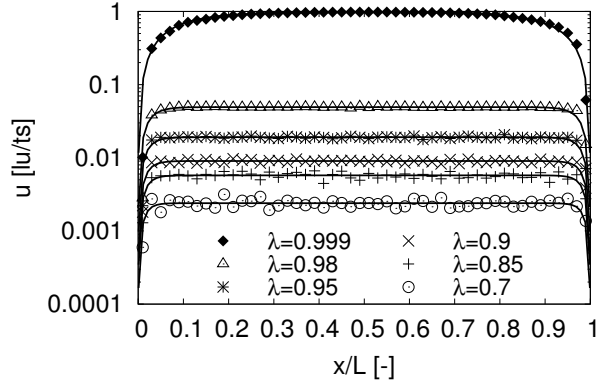}
\caption{Velocity profiles in the flow through a porous channel. Data points represent MPCD simulation, solid lines are fits to Eq.~\ref{eq:solution}.\label{fig:darcy}}
\end{figure}

Next, we investigated the relation between the two main parameters of the model: $r$ and $\alpha$ from Eq.~(\ref{eq:solution}). We used the data from the least square fits to the numerical profiles of the velocity. The test was repeated several times for varying $\lambda$. Numerical profiles of the velocity profiles with fitted analytical profiles are plotted in Fig.~\ref{fig:darcy}.
The numerical data agrees with the theoretical curves well. However, with the decreasing $\lambda$, a statistical noise appears. For example, at $\lambda=0.7$ some points at $x/L$ near $0.2$ and $0.65$ are laying relatively far from the theoretical profile. 
\begin{figure}[!ht]
\centering
\includegraphics[width=0.7\columnwidth]{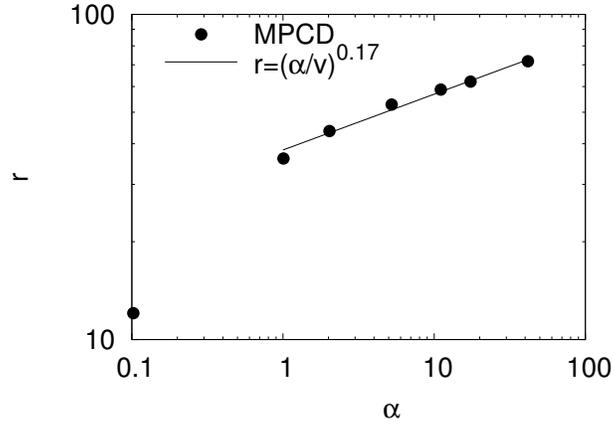}
\caption{Numerical test on the power law for the basic model parameters from Eq.~\ref{eq:solution}. The outlier at $\alpha=0.1$ corresponds to $\lambda=0.999$ which means highly permeable system (Poiseuille flow).\label{ralpha}}
\end{figure}
The values of $r$ and $\alpha$ corresponding to all velocity profiles measured in Fig.~\ref{fig:darcy} are plotted in Fig.~\ref{ralpha}.
We found that in our model the power law holds for $\alpha\geq 1$:
\begin{equation}
r \propto (\alpha / \nu)^{b},
\end{equation}
where $\nu$ is the model viscosity and $\alpha$ is the drag coefficient from Eq.~\ref{eq:solution}. The value $b\approx 0.17$ is smaller than $b=0.5$ found in a similar study based on the Lattice-Boltzmann model \cite{Dardis98}.

\begin{figure}[!ht]
\centering
\includegraphics[width=0.7\columnwidth]{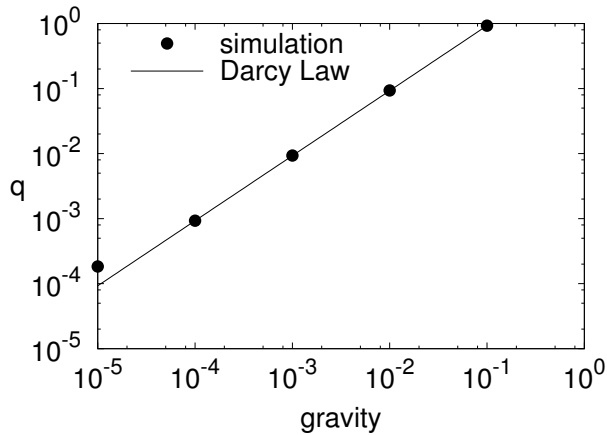}
\caption{Test of Darcy's Law $q\propto g$ - the dependency of the numerical flux $q$ on the gravity in the porous channel flow. $\lambda$ was fixed at $0.99$. \label{qg}}
\end{figure}

To study the permeability in our system we integrated the velocity along
the horizontal direction. The resulting plot of
the flux $q$ at varying gravity is expected to be linear and, in fact, linearity is seen in almost the whole range of gravity (see Fig.~\ref{qg}).

\begin{figure}[!ht]
\centering
\includegraphics[width=0.7\columnwidth]{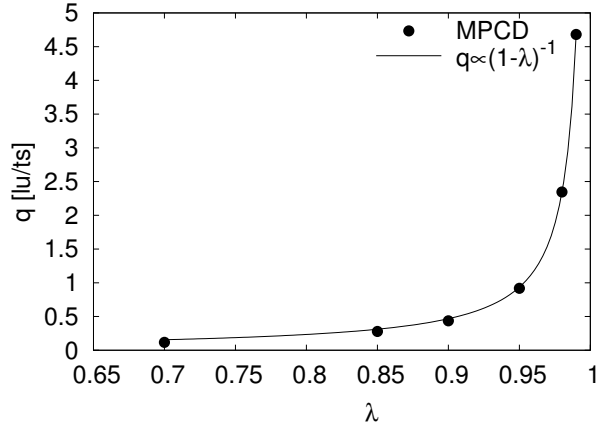}
\caption{Numerical fit of the numerical MPCD data to the function $k\propto 1/(1-\lambda$).\label{kns}}
\end{figure}

To measure the dependency of permeability on $\lambda$, we used directly the information hold in profiles in Fig.~\ref{fig:darcy}. We integrated the velocity in the profile to get the total flux $q$. We did this for each $\lambda$ and plotted the results in Fig.~\ref{kns}. 
Based on the results in Fig.~\ref{fig:darcy} and the previous observation that Darcy law holds in our system, we may now write that:
\begin{equation}
q\propto (1-\lambda)^{-1} \propto k.
\end{equation}

\section{Summary}

We extended the standard MPCD solver to account for porous media as a source of an additional force acting on a fluid. We confirmed that the extended model can reproduce the analytical solution to the damped Stokes model with the power law correlation between its two main parameters. We showed that the model correctly recovers the Darcy's Law. 
Our work has several interesting consequences. 
In principle, MPCD allows to include extra particles in the flow using molecular dynamics. Thus, the applications of our model may include other types of media, i.e. polymer flows, diluted particles transport, all in porous meda. Also, due to the possibility of modelling a spatial changes in permeability of the model by varying $\lambda$ in space, it is possible to simulate various types and shapes of porous structures, i.e. transport through rippled sea bed \cite{Huettel96,Kharab90}, layered porous media or porous particles \cite{Kindler10}.

%Our model may be also incorporated into other existing MPCD solvers and %extensions, i.e. ***, , the multiphase flows (***), free surface (***) or even %*** (*** do review). 
%\section{Results ideas}
%* shape of transition layer velocity (according to Reza)
%* transition layer for varying porosity (up to 0?)
%* flow over solid body (porosity 0?)
%* flow through porous particle (compare to Arzhang)
%* flow through porous ripples (compare to Hang Guo?) with photos from %Jannsand
%* Darcy profiles (like Dardis) - another article?

\section{Acknowledgment}

I would like to thank Zbigniew Koza, Arzhang Khalili and Jaros\l{}aw Go\l{}embiewski for reading the first version of the manuscript and providing an useful comments on it.

\section*{References}

\bibliography{tort}

\begin{thebibliography}{10}
\expandafter\ifx\csname url\endcsname\relax
  \def\url#1{\texttt{#1}}\fi
\expandafter\ifx\csname urlprefix\endcsname\relax\def\urlprefix{URL }\fi
\expandafter\ifx\csname href\endcsname\relax
  \def\href#1#2{#2} \def\path#1{#1}\fi

\bibitem{Huettel96}
M.~Huettel, W.~Ziebis, S.~Forster,
  \href{http://dx.doi.org/10.4319/lo.1996.41.2.0309}{Flow-induced uptake of
  particulate matter in permeable sediments}, Limnology and Oceanography 41~(2)
  (1996) 309--322.
\newblock \href {http://dx.doi.org/10.4319/lo.1996.41.2.0309}
  {\path{doi:10.4319/lo.1996.41.2.0309}}.
\newline\urlprefix\url{http://dx.doi.org/10.4319/lo.1996.41.2.0309}

\bibitem{Guo02}
Z.~Guo, T.~S. Zhao,
  \href{http://link.aps.org/doi/10.1103/PhysRevE.66.036304}{Lattice boltzmann
  model for incompressible flows through porous media}, Phys. Rev. E 66 (2002)
  036304.
\newblock \href {http://dx.doi.org/10.1103/PhysRevE.66.036304}
  {\path{doi:10.1103/PhysRevE.66.036304}}.
\newline\urlprefix\url{http://link.aps.org/doi/10.1103/PhysRevE.66.036304}

\bibitem{Gao12}
H.~Gao, M.~Matyka, B.~Liu, A.~Khalili, J.~E. Kostka, G.~Collins, S.~Jansen,
  M.~Holtappels, M.~M. Jensen, T.~H. Badewien, M.~Beck, M.~Grunwald,
  D.~de~Beer, G.~Lavik, M.~M.~M. Kuypers,
  \href{http://dx.doi.org/10.4319/lo.2012.57.1.0185}{Intensive and extensive
  nitrogen loss from intertidal permeable sediments of the wadden sea},
  Limnology and Oceanography 57~(1) (2012) 185--198.
\newblock \href {http://dx.doi.org/10.4319/lo.2012.57.1.0185}
  {\path{doi:10.4319/lo.2012.57.1.0185}}.
\newline\urlprefix\url{http://dx.doi.org/10.4319/lo.2012.57.1.0185}

\bibitem{Lin15}
D.~Lin, E.~Eek, A.~Oen, Y.-M. Cho, G.~Cornelissen, J.~Tommerdahl, R.~G. Luthy,
  Novel probe for in situ measurement of freely dissolved aqueous concentration
  profiles of hydrophobic organic contaminants at the sediment–water
  interface, Environmental Science \& Technology Letters 2~(11) (2015)
  320--324.
\newblock \href {http://dx.doi.org/10.1021/acs.estlett.5b00239}
  {\path{doi:10.1021/acs.estlett.5b00239}}.

\bibitem{Angelis12}
E.~De~Angelis, M.~Chinappi, G.~Graziani, Flow simulations with multi-particle
  collision dynamics, Meccanica 47~(8) (2012) 2069--2077.

\bibitem{Malevanets99}
A.~Malevanets, R.~Kapral,
  \href{http://scitation.aip.org/content/aip/journal/jcp/110/17/10.1063/1.478857}{Mesoscopic
  model for solvent dynamics}, The Journal of Chemical Physics 110~(17) (1999)
  8605--8613.
\newblock \href {http://dx.doi.org/http://dx.doi.org/10.1063/1.478857}
  {\path{doi:http://dx.doi.org/10.1063/1.478857}}.
\newline\urlprefix\url{http://scitation.aip.org/content/aip/journal/jcp/110/17/10.1063/1.478857}

\bibitem{Reid09}
D.~A.~P. Reid, H.~Hildenbrandt, J.~T. Padding, C.~K. Hemelrijk,
  \href{http://link.aps.org/doi/10.1103/PhysRevE.79.046313}{Flow around
  fishlike shapes studied using multiparticle collision dynamics}, Phys. Rev. E
  79 (2009) 046313.
\newblock \href {http://dx.doi.org/10.1103/PhysRevE.79.046313}
  {\path{doi:10.1103/PhysRevE.79.046313}}.
\newline\urlprefix\url{http://link.aps.org/doi/10.1103/PhysRevE.79.046313}

\bibitem{Bedkihal2013}
S.~Bedkihal, J.~C. Kumaradas, K.~Rohlf,
  \href{http://dx.doi.org/10.1007/s10237-012-0454-z}{Steady flow through a
  constricted cylinder by multiparticle collision dynamics}, Biomechanics and
  Modeling in Mechanobiology 12~(5) (2013) 929--939.
\newblock \href {http://dx.doi.org/10.1007/s10237-012-0454-z}
  {\path{doi:10.1007/s10237-012-0454-z}}.
\newline\urlprefix\url{http://dx.doi.org/10.1007/s10237-012-0454-z}

\bibitem{Bolintineanu12}
D.~S. Bolintineanu, J.~B. Lechman, S.~J. Plimpton, G.~S. Grest,
  \href{http://link.aps.org/doi/10.1103/PhysRevE.86.066703}{No-slip boundary
  conditions and forced flow in multiparticle collision dynamics}, Phys. Rev. E
  86 (2012) 066703.
\newblock \href {http://dx.doi.org/10.1103/PhysRevE.86.066703}
  {\path{doi:10.1103/PhysRevE.86.066703}}.
\newline\urlprefix\url{http://link.aps.org/doi/10.1103/PhysRevE.86.066703}

\bibitem{Wysocki10}
A.~Wysocki, C.~P. Royall, R.~G. Winkler, G.~Gompper, H.~Tanaka, A.~van
  Blaaderen, H.~Lowen, Multi-particle collision dynamics simulations of
  sedimenting colloidal dispersions in confinement, Faraday Discuss. 144 (2010)
  245--252.
\newblock \href {http://dx.doi.org/10.1039/B901640F}
  {\path{doi:10.1039/B901640F}}.

\bibitem{Dardis98}
O.~Dardis, J.~McCloskey,
  \href{http://link.aps.org/doi/10.1103/PhysRevE.57.4834}{Lattice boltzmann
  scheme with real numbered solid density for the simulation of flow in porous
  media}, Phys. Rev. E 57 (1998) 4834--4837.
\newblock \href {http://dx.doi.org/10.1103/PhysRevE.57.4834}
  {\path{doi:10.1103/PhysRevE.57.4834}}.
\newline\urlprefix\url{http://link.aps.org/doi/10.1103/PhysRevE.57.4834}

\bibitem{Balasubramanian87}
K.~Balasubramanian, F.~Hayot, W.~F. Saam,
  \href{http://link.aps.org/doi/10.1103/PhysRevA.36.2248}{Darcy's law from
  lattice-gas hydrodynamics}, Phys. Rev. A 36 (1987) 2248--2253.
\newblock \href {http://dx.doi.org/10.1103/PhysRevA.36.2248}
  {\path{doi:10.1103/PhysRevA.36.2248}}.
\newline\urlprefix\url{http://link.aps.org/doi/10.1103/PhysRevA.36.2248}

\bibitem{Lamura01}
{Lamura, A.}, {Gompper, G.}, {Ihle, T.}, {Kroll, D. M.},
  \href{http://dx.doi.org/10.1209/epl/i2001-00522-9}{Multi-particle collision
  dynamics: Flow around a circular and a square cylinder}, Europhys. Lett.
  56~(3) (2001) 319--325.
\newblock \href {http://dx.doi.org/10.1209/epl/i2001-00522-9}
  {\path{doi:10.1209/epl/i2001-00522-9}}.
\newline\urlprefix\url{http://dx.doi.org/10.1209/epl/i2001-00522-9}

\bibitem{Kharab90}
A.~Kharab,
  \href{http://www.sciencedirect.com/science/article/pii/0898122190900022}{A
  poiseuille flow past a permeable hill}, Computers \& Mathematics with
  Applications 19~(2) (1990) 9 -- 15.
\newblock \href
  {http://dx.doi.org/http://dx.doi.org/10.1016/0898-1221(90)90002-2}
  {\path{doi:http://dx.doi.org/10.1016/0898-1221(90)90002-2}}.
\newline\urlprefix\url{http://www.sciencedirect.com/science/article/pii/0898122190900022}

\bibitem{Kindler10}
K.~Kindler, A.~Khalili, R.~Stocker,
  \href{http://www.pnas.org/content/107/51/22163.abstract}{Diffusion-limited
  retention of porous particles at density interfaces}, Proceedings of the
  National Academy of Sciences 107~(51) (2010) 22163--22168.
\newblock \href
  {http://arxiv.org/abs/http://www.pnas.org/content/107/51/22163.full.pdf}
  {\path{arXiv:http://www.pnas.org/content/107/51/22163.full.pdf}}, \href
  {http://dx.doi.org/10.1073/pnas.1012319108}
  {\path{doi:10.1073/pnas.1012319108}}.
\newline\urlprefix\url{http://www.pnas.org/content/107/51/22163.abstract}

\end{thebibliography}

\end{document}